# Theory of cyborg: a new approach to fish locomotion control

Mohammad Jamali[1] , Yousef Jamali[2*], Mehdi Golshani[1,3]

**Abstract**:

Cyborg in the brain-machine interface field has attracted more attention in recent years. To control a creature via a machine called cyborg method, three stages are considerable: stimulation of neurons, neural response, and the behavioral reaction of the subject. Our main concern was to know how electrical stimulation induces neural activity and leads to a behavioral response. Additionally, we were interested to explore which type of electrical stimulation is optimal from different aspects such as maximum response with minimum induction stimulus field, minimum damage of the tissue and the electrode, reduction of the noxiousness of stimuli or pain in the living creature. In this article, we proposed a new model for the induction of neural activity led to locomotion responses through an electrical stimulation. Furthermore, based on this model, we developed a new approach of electrical neural stimulation to provide a better locomotion control of living beings. This approach was verified through the empirical data of fish cyborg. We stimulated the fish brain by use of an ultra-high frequency signal which careered by a random low frequency. According to our model, we could control the locomotion of fish in a novel and innovative way. In this study, we categorized the different cyborg methods based on the nervous system areas and the stimulation signal properties to reach the better and optimal behavioral control of creature. According to this, we proposed a new stimulation method theoretically and confirmed it experimentally.

1  Department of Physics, Sharif University of Technology, Tehran, Iran

2  Department of Applied Mathematics, School of Mathematical Sciences, Tarbiat Modares University, Tehran, Iran

3  School of Physics, Institute for Research in Fundamental Sciences (IPM), Tehran, Iran

\* Correspondence: y.jamali@ipm.ir

# Introduction

"Cyborg" combined the words "cybernetic" and "organism.". one of the definitions of this concept refers to a combination of machine and living organism, devices that enable to conduct the organism behavior especially its movement. In this context, cyborg has attracted growing attention in the field of brain-machine interface. The locomotion control of living creatures, in its modern form, was initiated in 1997 by using cockroach, which has simple neural structure [1]. However, the control of the cockroach locomotion is still under investigation [2-4]. Generally, most of the studies have focused on the cyborg insects, such as beetles and moth [3, 5-15]. Usually, this aim is achieved by electrical stimulation of the nervous system including stimulation of the afferent nerves, the efferent nerves which especially are related to the flying muscles, and recently, via ganglions stimulation [3]. In order to understand the brain structure and motor locomotion processes, in the last 60 years, many studies have been done on the neural system of aquatics [16-21] and mammals [22], through electrical stimulation method. Recently, significant studies have been done on the locomotion control of rat [23-25], reptiles [26, 27], and on the control of the pigeon's locomotion and flight [28-31].

The first report on locomotion control of fish was published in 1974 by Kashin [17]. These researches were continued until 2000 in an effort to understand the mechanism of neural function in the locomotion of aquatic species, including lobster[16], rays[32-34], teleost fish (carp [20] and redfish [35]), and eel [36]. Some scientific reports on the cyborg fish (bio-robot) were published in the last decade, tried to control the fish locomotion for the right, left and front in the free swimming[18, 20]. During 2009 to 2013, Similar work was done via stimulation of the midbrain locomotors nucleus. These researches were done on the controlling of the locomotion of redfish and crap [18, 19, 21] by focusing on the design of an effective implanted electrode.

The main method of the cyborg is the electrical stimulation of its neural system or muscles. In the cyborg stimulation, the usual method is applying an electric field to the implanted metal electrode, as is the case of deep brain stimulation (DBS) and electrical brain stimulation (EBS). An electrical stimulation in the cyborg must satisfy some conditions, like as locality [7], uniqueness of response to the same stimulation, minimal damage to the tissue and electrode, no reduction of the lifespan in the natural conditions, reduction of consumption power of stimulator device, and reduction of the creature pain (persecution).

This article contains a theoretical part and an empirical one, which includes a review of cyborg theory and suggests a new sight into stimulation. Ultimately, we indicate a new method of neural system stimulation, which results in a novel method of fish locomotion control. We control the fish locomotion to the left, right and front in free swimming, using a very low charge injection with minimal damage to the tissue.

**Theoretical part**

- Theory of cyborg's motion control

All living creatures, from simple organisms like bacteria to human as most complex one, [37, 38], for efficient active movement need some tools to sense environmental signals, analyze them, and react accordingly (respond to them). Interestingly, there is some claim that during evolution, the nervous systems and brain evolved to control movement [39]. In the phylogenetic tree of life, as we close to the primitive organisms, their neuronal structure becomes simpler and this network exhibits more linear response to sensory stimuli and therefore the neural response to the particular sensory input is more predictable (Even in human, many of his/her nervous system's responses to sensory stimuli are automatic and predictable.). Therefore, in the controlling of a cyborg, where the aim is not to train the creature but to control its motion (i.e. making the decision instead of its nervous system), controllability of simplest creature is more feasible.

In general, there are two approaches in the neural stimulation used for controlling the locomotion of living creature: one of them causes a disorder or disability through the disturbing the dynamics of nervous system (as is done in [3, 5]); another one, we termed active stimulation, forms an appropriate response by triggering neural impulses. There are three types of active stimulation based on the targeted neural network: a) stimulation of afferent (sensory) nerves (e.g. [1, 3]), b) stimulation of efferent (motor) nerves (e.g. [6]), and c) stimulation of interneurons (brain or ganglion) that is related to the decision making (e.g. [5]).

There are two types of sensory stimulation: The first is the stimulation with a short duration time which causes a short-time locomotion response and for example is used in direction shifting (turning) [5, 6], or initiation of a neural activity processes in the brain, like as start of flying [5]. In the last example, the activity usually begins at the moment of cutting off the stimulation. The second is a long-term stimulation that causes a corresponding long-term locomotion response, like a lateral line sense stimulation of fish for making a forward movement [17]. The second type is more used for senses such as hearing or location searching through the antenna [1, 2, 40]

In simple creatures (i.e. creatures with simple nervous system which not develop as much like as lower invertebrates), because of the linear reaction to the sensory stimulus, one can control them by stimulating their sensory nerves. However, due to the existence of other senses and the fact that all of the signals from them are analyzed in the brain (the ganglion) simultaneously, precise locomotion control is not realizable. In other words, other natural sensory inputs may interfere and reduce the effect of stimulation [5].

It seems that in the natural conditions if the environmental signal for a sensory system is very simple and more localizable in the frequency space, the feasibility of stimulation with the conventional methods (i.e. by electrode implanting and applying a simple alternating electric current to it), are more likely. Therefore, the easiest way for controlling is finding and using the sensory system with simple input. Some examples of this type of sensory system are auditory system, cockroach cerci, cockroach antenna, and fish Mauthner cells. Similarly, the more sense input data are abstract (i.e. need lower analysis and processing by the central nervous system) the more their effect is linear on the locomotion output.

This remains valid even for the brain, i.e. the stimulation of lower levels of the brain which have a more specific frequency range[41], is more effective.

In addition, in our opinion, the coordinated stimulation of different sensory neural areas causes a more predictable neural response. Hence, in this circumstance a more linear stimulus-response behavior is achievable. It is because the neural system tries to eliminate the waste information of the variety of sensory input data and to send more coordinated information to the central neural system. The difference between the neural pattern of electrical stimulation and natural stimulation confirms this idea [42]. Hence, we propose that it is possible to achieve more determinable response by a consistent and coordinated stimulation of different sensory systems (e.g. olfactory, hearing, and tactile).

The advantages of the afferent nerve stimulation are less damage to the neural tissue and no disturbing of the complex brain processes. Indeed, the stimulation of sensory nerves, which have a short frequency bandwidth [41], is easier, and response of the central neural system to it will be more linear. In those nerves which pass a wide range of frequencies (such as sensory nerve in comparison to pain or visual relative to hearing) [41], more complex electrical stimulation is needed for effective stimulation.

The next stimulation type is the stimulation of efferent nerves. This was used in [5-7]. The higher certainty of response to stimulation is one of its advantages. But because of the complexity of locomotion process and a large number of involved muscles, the controlling of this collection is very difficult and inefficient. However, when the aim is to make a small disturbance in primary locomotion, this type is very appropriate. For example, in beetle free-flying, the third basilar muscle causes a turning by producing an imbalance motion between the two wings. Therefore, the stimulation of this muscle can cause the beetle turns in free-flying [6].

The third type of stimulation is the central nervous system (brain or neural ganglions) stimulation which its mechanism is still very crude and complex. The stimulation of the brain stem, specially midbrain, to excite motor nucleus was done in [16-22, 26, 27, 30, 32-36, 43], and the stimulation of the regions of the cerebellum was reported in [44]. The stimulation of ganglion was used in cockroach control [3] and in our lab was done on the locust and dragonfly to control their motion during fly.

It seems that, in the first step, those types of movements that have an alternating and rhythmic pattern, such as insect locomotion, forward swimming in the fish and insect wing fluttering are more suitable for control by stimulation. Because of the alternating pattern of these types of motions, the corresponding motor nervous system mostly responds better to an alternating stimulation signal. In other words, a periodic stimulation can force the neural system to send an appropriate periodic signal for generating these types of motions (such as motor nucleus stimulation in fish and cat [17, 22]). In addition, disturbance of this neural system can cause a disorder in the normal alternative signal which in turn, forms a new response like as turning [3].

The experimental results show that the response depends on the frequency range of stimulation. For example, our experiments show that in the forward control of fish locomotion or in the locust flying control, the nerve system only responds to a very narrow range of stimulation frequency (respectively between 70-90 Hz and 320-350 Hz).

The activity of the neurons of the motor cortex, responsible for locomotion, is in the limited frequency range, according to their simple and specific tasks [41] and therefore, their stimulation is easier. But the stimulation of some parts of deeper regions of the brain, which are responsible for processing and decision making, to the aim of locomotion control is challenging, due to three reasons: 1. Because of their wide functional and structural properties (i.e. their complex task, type diversity, and complex physical shape). Hence, these neurons have a wide bandwidth frequency [41]. 2. There are a large number of connections between different areas and many inhibitory neurons involved to control these connections and activity. It can lead to a lack of unique response. In addition, during the frequent stimulation, the excessive excitation of inhibitory neurons sometimes causes the disturbing of the neurotransmitters balance. It seems that the report on the combination of rotary fish motion with the forward movement in [35], by cerebral stimulation, is due to this fact. Our experiment on deeper regions of fish cerebellum confirms this phenomenon too. 3. There is no one-to-one and direct relation between these regions and output motions. Deep areas of the brain are associated with different parts of the upper areas, which are responsible for various muscles control.

Thus, we could conclude that in the sensory nerve stimulation, the frequency of stimulation must be localized, as it causes the reduction of input signal entropy (i.e. the increase of information). In contrast, at the DBS, having a higher entropy to use the whole frequency bandwidth capacity of neurons is recommended especially in low intense stimulation. In addition, the deep brain stimulation should be more localized in space to avoid the excitation of other regions due to large connections between neighboring areas.

- **Optimal stimulation**

As we mentioned above, an optimal electrical stimulation in cyborg should satisfy the following propositions:
1. Maximum response with minimum induction stimulus field
2. Minimum damage of the tissue and the electrode
3. Reduction of the noxiousness of stimuli or pain in the living creature involved
4. Increasing the battery lifetime

Two important electrochemical phenomena can be observed during applying voltage or injecting an electric current to the tissue via an implanted metal electrode:

1. The reorientation of bipolar water molecules and the formation of an ionic layer around the electrode create a double layer capacitance. Some parts of the injection charge are stored in this double layer capacitor, whose capacitance depends on the shape and material of electrode [45, 46].
2. If the time duration of direct stimulation is long enough, then, according to Butler-Volmer equation[46], after charging of double layer capacitor, a direct current (DC), named Faradaic current [45, 47], flows. In this current, there is the exchange of electrons between the electrode and the tissue [48]. The associated impedance of this current, in the final equilibrium state, is about of few megaohms.

The Faradaic current causes tissue damage, electrode corrosion, production of toxins in the tissue and creation of gas bubbles at the electrode surface [46, 47]. In EBS, the reduction of the Faradaic current is one of the main issues.

Some of the electrochemical effects spread out in the tissue due to the diffusion of ions in the neural tissue. This process is irreversible in prolonged direct stimulation. At the enough short time stimulation, since the electrochemical product does not move away from the electrode, by inverting the direction of current/voltage, some product that has been recently formed may be reversed back into its initial form (e.g. the ions cannot get too far away from the electrode, they return and gather in the inverse step). In this condition, the electrochemical effects can be reversible [46, 48]. In other words, at inverse physical conditions (charge, current or voltage), these effects are canceled, and this ensures the minimally of destructive effects [47].

Regardless of the frequency of the stimulation, there are two types of stimulation signals in the Cyborg and EBS: 1. Stimulation with unipolar waveform: in which periodic positive (negative) pulses, are applied to the tissue with a certain time duration (Figure 1a). 2. Stimulation with bipolar waveform: this stimulation contains periodical and symmetrical waves composed of positive and negative pulses (Figure 1b-d).

In principle, stimulation with higher frequencies results in the reduction of electrochemical effects [47-54], especially when the induction current has the charge balance (i.e. pure injected charge at each period is zero or minimum) [47-52]. Hence, in order to produce a stimulation with the minimum damage of the tissue and electrode, the induced current should be with charge balance and its Faradaic part should be minimal. Figure 1 presents different types of stimulation signals based on the mentioned features.

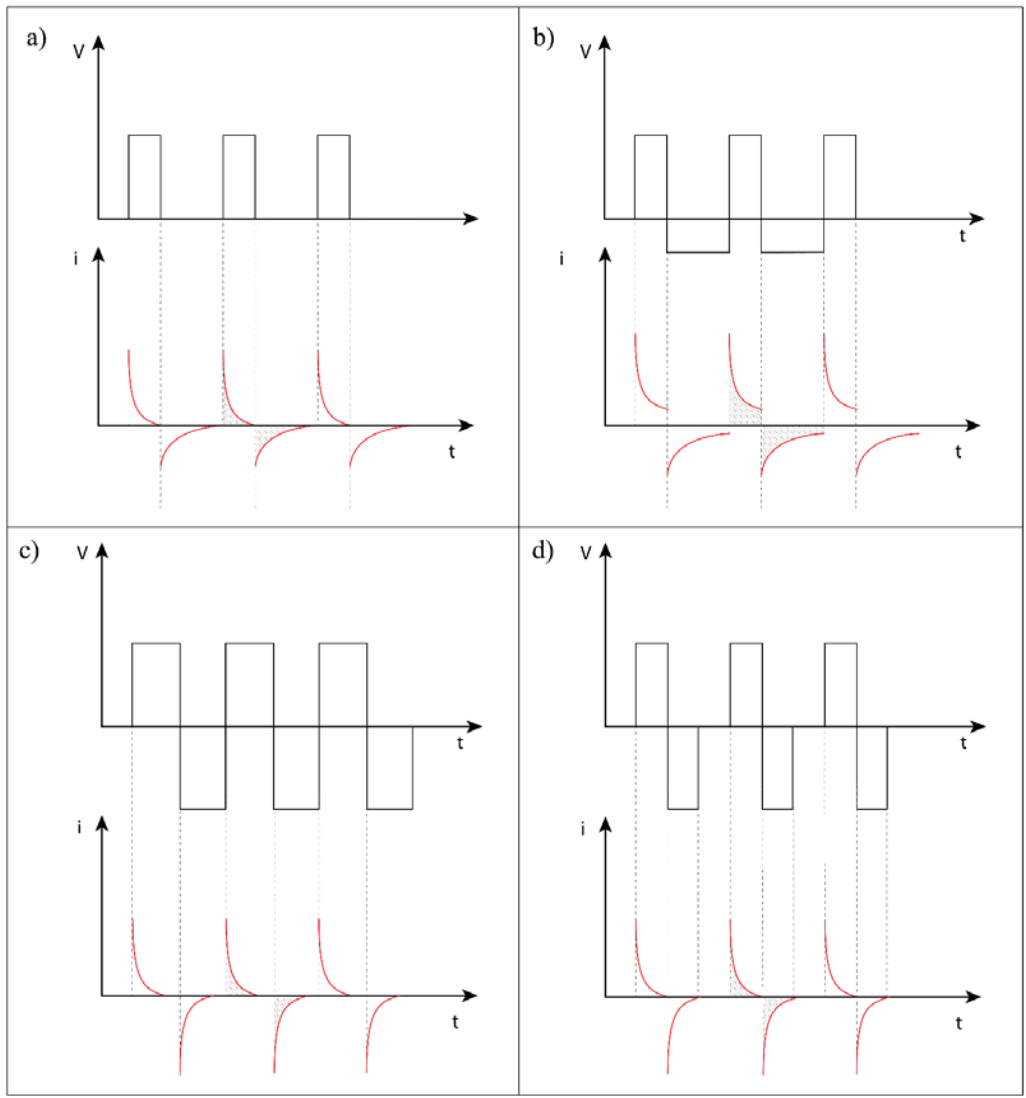

*Figure 1. several types of charge balance stimulation signal. Charge balance refers to the balance of the injection charge through the stimulation. It means that at any period of stimulation the amount of charge injected to tissue and amount of charge which returns to the electrode are equal. Amount of injection charge is equal to the area under the current curve. At all pictures, we have the charge balanced.*

The selected shape of the stimulation signal depends on the tissue and the neuronal system properties. The physical properties of tissue determine the frequency and the intensity of the stimulation. In fact, the property of neural system does not allow to increase the stimulation frequency arbitrary. Hence, to reduce Faradaic current, one can insert a capacitor in the path of the electrode whose value is around the double layer capacitance (Figure 2).

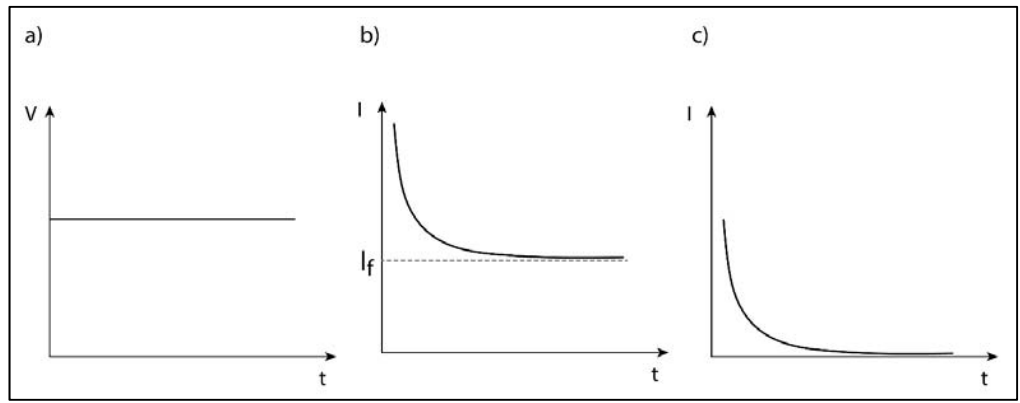

*Figure 2. Schematic of constant-voltage stimulation and its correspondence induced current in tissue. A. Voltage versus time graph in a constant-voltage stimulation. B. Induced current diagram without the inserted capacitor. C. Induced current at the presence of a capacitor in the path of the electrode, the main part of Faradaic current is removed at the effective time of stimulation.*

The Faradaic current causes the corrosion and oxidation of the electrode and damage to the tissue (Figure 3).

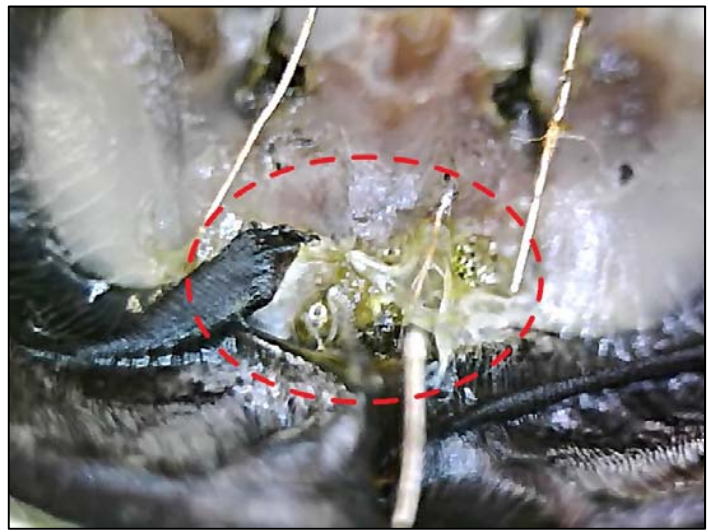

*Figure 3. central ganglia and optic lobs in the dragonfly. Oxidation of electrodes (green area) due to the Faradaic current cause permanent damage in tissue and has a toxic effect on tissue.*

In the electrical stimulation, the type of electrode is very important in the value of charge injection and in the toxic effect. The choice of an electrode, with higher capacitance, can make the stimulation less destructive. Table 1 and 2 in ref [47] summarize a list of electrodes in terms of their toxicity and their capacitance value.

Thus, it is impossible to propose a unique protocol for non-destructive stimulation of all areas of the neural system. The optimal stimulation is a function of a variety of parameters, some of which are out of control, and depend on the properties and characteristics of the tissue. Nevertheless, to optimize the stimulation, one can use a suitable capacitor and the electrodes with high capacitance and use symmetric stimulation signal with charge balance. As mentioned, in addition to the aforementioned methods of control, in some cases, we need to eliminate the function of the nerve activity or disturb it to affect the subject motion, as were done on beetle fly's stopping [5, 7] or cockroach's turning [3]. A few methods are available for neuron blocking:

1. Applying a long-term DC pulse: The production of a long-term DC signal through applying a constant current will cause the inactivation of sodium channels [47] which, in turn, causes neuron blocking. This method has been used in cyborg beetle [5]. In addition to the increase of the probability of tissue destruction, one of the problems of this method is that after stopping the stimulation, an intense neural excitation can be generated [47, 55].
2. Applying an alternatively high-frequency signal with a long-term duration: This high-frequency signal ($\sim 500 Hz$) results in the neural fatigue effect, and after the beginning of stimulation, the number of neurotransmitters began to decrease. The lack of neurotransmitters causes to disconnection of neurons [56-61]. The loss of neurotransmitter balance, particularly in the brain, can have negative long-term effects on the nervous system's performance.
3. The Production of a pulse or alternating signal with high amplitude: This highly intense stimulation causes axons' hyperpolarization, and will inactivate them [45].
4. The Production of a very high frequency ($\sim 10 KHz$) signal with high intensity: This signal can cause the blocking of neuron conductivity [56], and due to its higher frequency relative to the previous one, it results in less damage to the tissue. In this method, at the beginning of stimulation and before the neuron blocking, a highly intense "onset response" is observed [62-67], which could be avoided by the methods proposed in reference [56].
5. The production of a very high-frequency signal with low amplitude: This signal causes some perturbation in the rate of sending information. After a very short reaction, the neuron and neural system are adapted to it.

It seems that due to the adverse electrochemical effects of DC signal which is due to the Faradaic current creation, the first method can only be used in a limited way. Although other methods are common in DBS and EBS, to the best of our knowledge, nothing has been reported on the cyborg. However, because of their less destructive effects, we recommend using them in the cyborg field.

- **Locality in stimulation**

Localization and limitation of the stimulation to a finite group of neurons are very important to control the cyborg. Applying an alternating electric voltage to the electrode induces a quasi-stationary or quasi-static [68] current, in the ionic medium. The induced electric field or its equivalent electric current ultimately will initiate neural firing process, by opening the sodium channels. What determines the degree of stimulation locality is the diameter of the implanted electrode and the intensity of stimulation[69]. The following relations show the shape of the induced electric field (E) in an ionic environment (with a cylindrical symmetry):

$$|E| = \frac{I_0 \omega}{2\pi\sqrt{\sigma^2 + \epsilon^2 \omega^2} R} + \frac{I_0 l \omega}{2\pi\sqrt{\sigma^2 + \epsilon^2 \omega^2} R^3}$$

$$|E| = I_0 \omega \left(\frac{\alpha}{R} + \frac{\beta}{R^3}\right)$$

Where $I_0$ is the intensity (amplitude) of the current source, ω is the angular momentum of alternating current, σ is the conductivity coefficient, ϵ is the absolute permittivity of the ionic environment, and R is the vertical distance to the electrode axis. Furthermore, α and β are two positive constants.

These relations indicate that the intensity of stimulation will be decay fast as the distance is increased. In the cyborg case, where the size of the electrode diameter is usually around 70 to 200 micrometer and its shape is cylindrical, the stimulation is highly localized. This mathematical formalism is confirmed by some experimental data [69]. For electrical stimulation, two electrodes as positive electrode and ground electrode are used. Although in most cases, the distance between these two electrodes is large enough, (as will be explained below,) it seems that stimulation is sufficiently localized, and farther points are not directly affected by stimulation. In addition to the above explanation, the shielding effect of ions and bipolar water molecules decrease the field effect dramatically at further distances. In fact, when electrodes are away from each other, the induced electric field near one of them is not affected by the other one and has circularly or cylindrical symmetry due to the shape of the electrode. But by bringing the electrodes near each other, the effect of ground electrode become important and the electric field will be oriented from positive to the ground electrode with a higher intensity. Therefore, decreasing the distance between electrodes cannot improve locality of stimulation, even it can cause high polarization of ionic tissue and the reduction of equivalence resistance and the increasing of Faradaic current [48]. However, more localized stimulation can be achieved by designing an array of micro-electrodes to concentrate the induced field into a small region and eliminate it in other regions [32, 33].

- **Increasing the stimulation probability by increasing the frequency**

To increase the probability of neuron response to stimulation, two ways can be proposed: 1. increasing the injected charge and/or 2. increasing the time duration of stimulation by increasing the frequency of the signal[4]. The main privilege of the latter way is that there are no adverse effects of increasing the electric current and charge injection. By using this method, it is possible to increase the neural response probability via a weak stimulation. Table.1 shows the amount of charge injection under the different stimulation frequencies in the fish locomotion control experiment. However, stimulation by an ultra-high frequency leads to the fast neural system/neuron adaptation and no longer response [56-61]. To overcome this challenge in the cyborg cases, we suggest modulating the ultra-high frequency ($\sim 100 KHz$) by a low frequency carrier signal ($\sim 100 Hz$) which causes a more optimized stimulation with less current intensity, lower charge injection, less battery consumption and so less damage to the tissue.

*Table 1. Table of stimulation signal properties contain the type of stimulation (random or regulate), the main frequency, carrier frequency, duty and injected charge at each half of stimulation period.*

| Type of stimulation | On & off Mean frequency (Hz) | Frequency of stimulation | Duty (%) | Charge amplitude (nC) |
|---|---|---|---|---|
| Random | 200 | 1 MHz | 5 | 0.3 |
| Random | 200 | 200KHz | 5 | 3 |
| Regulate | 200 | 3.5KHz | 5 | 30 |
| Regulate | 150 | 150Hz | 5 | 60 |

Based on our theoretical and computational studies [unpublished work], we suggested that for producing a stimulation with low intensity and less damage, and for preventing the neural adaptation and optimal using of the information transferring capacity of neurons, the use of random low-frequency carrier signal is very effective (Figure 4). By a similar method, in the cyborg pigeons, the stimulation signal is a high-frequency signal (2 kHz), which is turning on and off at the frequency of 100Hz [28, 29].

---

[4] In fact, the effective time of stimulation is not equal to the time of stimulation at each stimulation signal period. When a pulse applies to the tissue, it is stimulated in the beginning of pulse duration, especially when we use the capacitor to decrease the faradaic current (see Figure3)

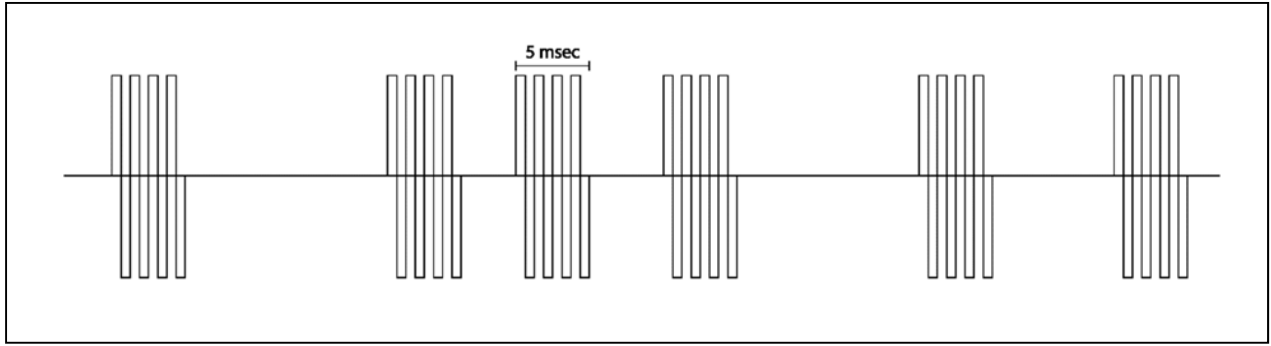

*Figure 4. Schematic of an ultra-high frequency stimulation signal with a random low-frequency carrier.*

- **Reduction of the noxiousness of stimuli**

One of the concerns in the cyborg scope is that in the stimulation of the sensory nerves in a certain region, there may be the likelihood of pain-sensing and/or persecution in the living creature. At the same signal duration of stimulation, there are a different required signal threshold between stimulation of a pain neuron and another type of sensory neuron [57, 70, 71]. Hence, it can be claimed that in the ultra-high frequency stimulation, by producing a proper signal frequency, with low amplitude, one can only stimulate sensory neurons without activation of the pain neurons. This difference, which is probably related to the difference of neurons' diameter, provides a cyborg stimulation protocol to reduce the amount of pain and suffering of the living creature.

- **Conclusion:**

Thus, to stimulate a neural system, our suggestion is the usage of a stimulation signal which has the following properties: a. An ultra-high frequency signal which turns on and off randomly at low frequency. b. The mean frequency of this random turning on/ off frequency is of the order of physiological frequency of the neural system. c. The shape of the high-frequency signal is bipolar with the charge balance. This method of reducing adverse electrochemical effects and power consumption causes an optimum stimulation with no (or minimal) pain and nuisance sensing for the creature. In addition, due to the weak intensity of signal power, this method like a natural stimulation causes some disturbance to the decision inputs of the living creature. It can be a new method for locomotion control of living creatures, especially for civilian purposes and bio-robotic network structures.

We used this method for the control of fish locomotion, as we shall describe in the following section.

# Experimental part

All experimental procedures were in accordance with the guidelines of the National Institutes of Health and the Iranian Society for Physiology and Pharmacology and were approved by the animal care and use committee of the Institute for Research in Fundamental Sciences.

- **Fish locomotion control**

Our sample was an African catfish. This aquatic vertebrate was selected for testing because of its special features, like high compatibility with the surroundings, the power of recovery and self-curing, low stress, and high survival ability outside water. This type of fish is usually active at nights work and has poor visual sight, and also two tentacles for environment searching through touching and short-range hearing.

As we mentioned earlier, there are three methods of stimulation with the aim of locomotion control. In the first method, which is the stimulation of afferent nerves, the two short range and long range hearing senses are good candidates for stimulation. The optic lobes and some areas of the cerebellum are proper candidates for the brain stimulation. As we mentioned earlier, cerebellum cortex is more suitable for fish locomotion control, because of its close connection to motor outputs and specific functional operation. However, the midbrain stimulation for locomotion control of fish has been used in [16-21]. Fish has two hearing senses. The main auditory information is received through outer and inner ears and entered to the hindbrain via VIIIth Cranial nerves (Figure 5) [72, 73].

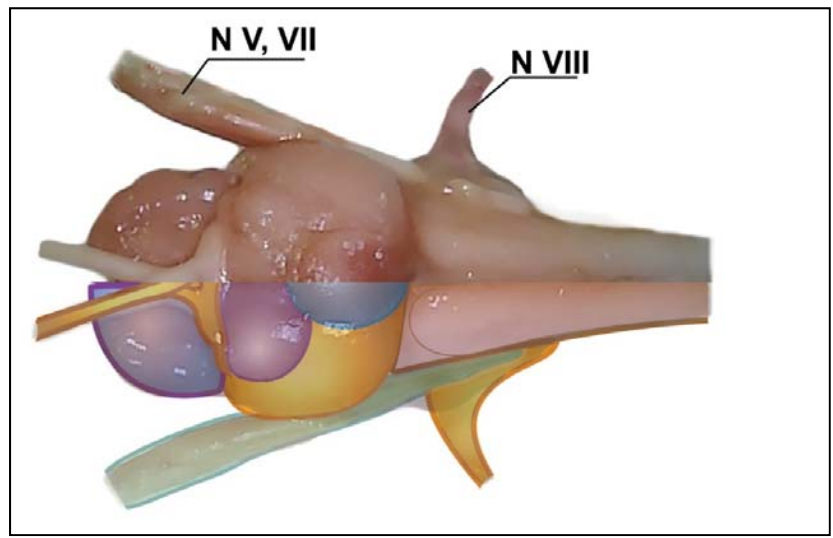

*Figure 5. The brain of an African catfish. The information of auditory sensing is entered to hindbrain via the VIIIth Cranial nerves. The information of sensory receptor of the head (especially the information of the anterior lateral line nerve) is entered to the brain through the VIIth and VIIth Cranial nerves.*

All points of the mentioned nervous system can be used for stimulation. However, due to the neuronal accumulation, especially the existence of nerves which are related to the involuntary actions like as heart rate [73, 74], electrical stimulation is not recommended by the conventional method. Hence, we stimulate auditory nerves and inner ear, which results in rotating the fish by creating a sense of sound. But experimentally we don't find it optimal, as compared with the stimulation of the short-range auditory sense. Fish senses surrounding short range flow and current via the lateral line on the fish body and the sensory receptors on its head [72, 75].

The lateral line information is entered into the brain via nervus linea lateralis posterior (nllp) from the behind of the brain, and the information of head receptors [73] is entered into the brain from the front via nervus linea lateralis anterior (nlla). The stimulation of short-range auditory sense via nllp nerve causes more effective stimulation for turning control of fish locomotion. We try this through three signals: with a low frequency (around 140 Hz), with an ultra-high frequency ($\sim KHz$) which turning on and off by frequency of 100 Hz, and with an ultra-high frequency signal which turned on and off randomly by a frequency in the range 100 to 200 Hz and with a Poisson distribution. All stimulation signals caused the turning of the fish to the opposite side. Discrete and in place turning were the topological property of respond to this stimulation.

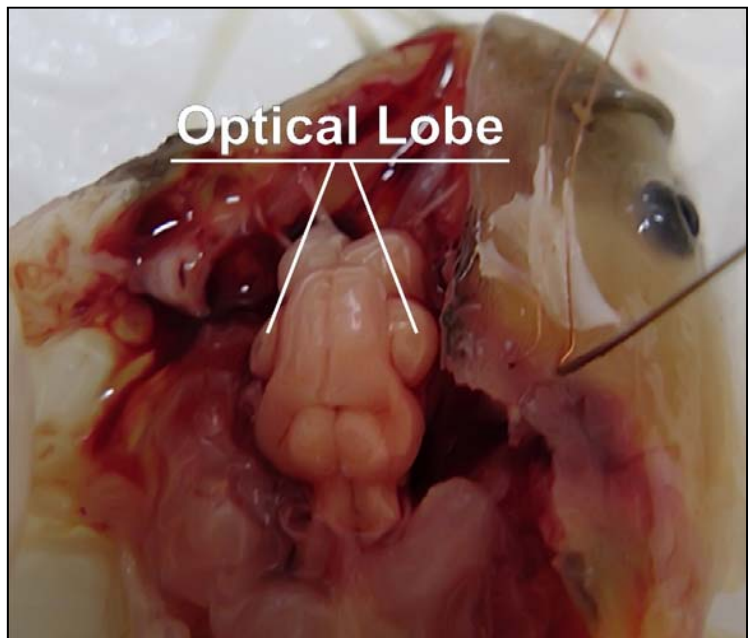

*Figure 6. optical lobes of African catfish.*

In addition, in the brain stimulation through the optic lob, we achieved the turning of fish to both sides (Figure 6). We observed the high sensitivity of response to the stimulation intensity. Therefore, it seems that auditory stimulation is more optimal than the optic lobs stimulation. According to what we mentioned in the previous section, this may be due to the complexity of visual information processing as compared with the auditory information.

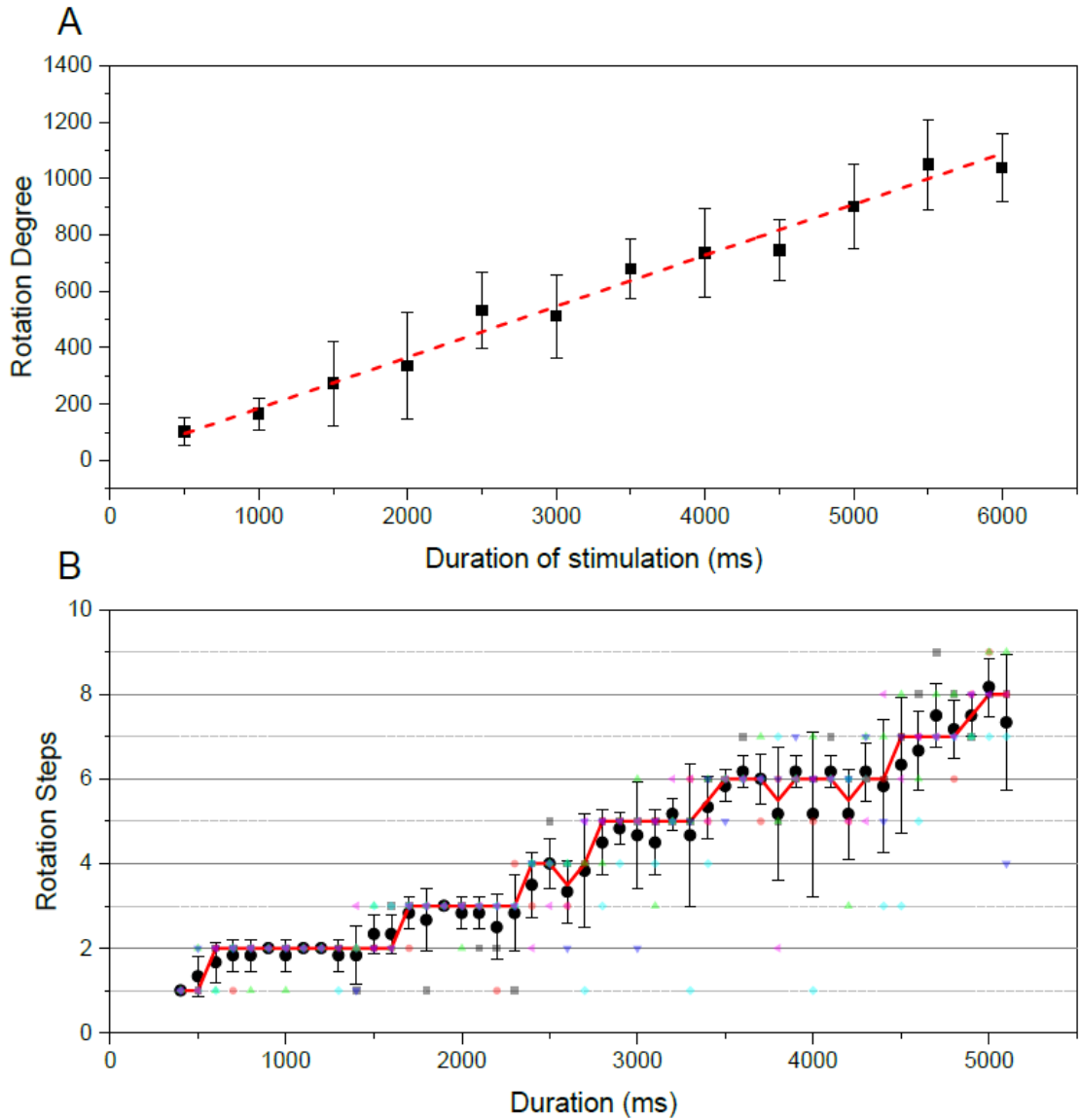

*Figure 7. Rotation of fish versus the time duration of stimulation. This experiment performed on the three samples of African catfish, and for each duration of stimulation, the experiment is repeated ten times, with a 30 s resting interval between each stimulation. A. Diagram of mean degree of rotation versus the time duration of the stimulation. By increasing time duration of the stimulation signal, the amount of rotation increase linearly. B. Due to the stepping structure of fish rotation (discrete steps of rotation), it is the diagram of the number of steps of rotation versus the time duration of stimulation. It is shown that the number of rotation steps is increase discreetly by increasing of the time duration of the stimulation signal.*

For forward locomotion control, the stimulation of the cerebellum anterior is appropriate. We stimulate this region by two types of stimulation signals: one with a low-frequency signal (= 80 Hz) and another with a high-frequency signal which randomly turns off and on with a mean frequency of 100 Hz. As theoretically expected, it is observed that the intense stimulation of the deeper areas results in a combination of two types of movement during the stimulation (rotation in line with the forward swimming) and interestingly this mixture movement persist for a couple of hours after stimulation experiments. As mentioned before, this may be due to the elimination of neurotransmitter balance and the perturbation of the inhibitory neurons normal activity. As was reported in [44], along with the forward locomotion of fish, a spiral motion around the fish longitudinal axis was observed. In other words, during this behavior, the stimulation of motor nucleus to forwarding locomotion results in the excitation of the motor nucleus which is responsible for the fish tail rotation. Due to the aerodynamic property of its water environment, this tail rotation causes a spiral motion around the longitudinal axis (due to the angular momentum conservation).

The stimulus of the left and right sides of the cerebellum, caused the fish to turn to the opposite sides. Topologically speaking, this turning was not in its place and was accompanied by forwarding locomotion, which was due to the symmetrical motion of the tail. Increasing the stimulation signal frequency caused an increase in the forward swimming velocity. In addition, stimulation by a low-intensity high-frequency signal caused smooth forward locomotion so that the fish could easily change its path to deal with obstacles. Despite the fact that a low-frequency stimulation causes fish to feel fatigue and lethargy after numerous and consistent stimulation, more the natural behavior without sensible fatigue was seen in stimulating by ultra-high frequencies signal. Therefore, it seems that this method would be better and more suitable for cyborg cases.

- **Material and method**

Stimulation was performed with a stainless (or gold plated copper wire) electrode with a diameter of 120 and 100 $\mu m$, with a cylindrical shape and the length of 1 $mm$. To reduce the Faradaic current, a 50 $\mu F$ capacitor was used in series. It is observed that the presence of this capacitor led to a better response qualitatively and lower power consumption. The stimulation signals were generated by a function generator through the voltage constant stimulation method. The turning signal on and off randomly via a handwriting code to control and change the signal parameters. The locomotion of fish is recorded by a camera. The average size of the fishes is 12 cm.

- **Empirical results**
1. Turning control of fish at low frequency (140Hz): Figure 7A shows the amount of rotation as a function of the time duration of stimulation. The fish rotation has a discrete form and is not continues, and Figure 7B shows the number of rotations as a function of the time duration of stimulation. The mean of RMS current is **50** microampere and the mean of injected charge per cycle is **0.25** microcoulomb. There is no evidence of fish nuisance, although, after a long period of stimulation (one hour), fatigue and anomalies are appeared in the fish behavior, which is disappeared after a while (some rest).

2. Rotation of fish at the ultra-high frequency randomly turned on and off from Poisson distribution by mean of 200Hz. Although increasing the signal frequency decreases the injection charge, it leads to the neural adaption effect. This is partially improved by turning on and off of the high-frequency signal by a low frequency, and it was greatly improved by the use of a random low frequency, with the Poisson distribution. The schematic of the applied signal is shown in Figure 4. There was no effect of nuisance and the fish behavior was very smooth and natural. Table.2 shows the amount of injected charge in each period, at the different stimulation signal frequencies. (supplementary movie 1)

*Table 2. The amount of injected charge in each half period, at the different stimulation signal frequencies.*

| Type of stimulation | On & off Mean frequency (Hz) | Frequency of stimulation | Duty (%) | Charge amplitude (nc) |
|---|---|---|---|---|
| Random | 200 | 300KHz | 15 | 0.7 |
| random | 200 | 2.5KHz | 15 | 12 |
| regulate | 150 | 150Hz | 15 | 140 |

3. Forward locomotion control of fish took place by the use of the signal frequency of 80 Hz and by using a series capacitor. The increasing of the signal frequency caused the increasing of the swimming velocity. The fish locomotion is natural.
4. Forward locomotion of fish with an ultra-high frequency signal, repeatedly turned on and off with 0.01 $ms$ intervals (low frequency), caused a very smooth and normal motion.

Supplementary movie 2 shows the locomotion of fish at three directions (left, right and forward) at an ultra-high frequency signal repeatedly turned on and off with 0.01 $ms$ intervals.

## Conclusion

In this article, a qualitative model of stimulation and its parameters (in terms of stimulation regions and type of stimulation signal), is presented with the aim of a locomotion control. In addition, the uniqueness of response, the mechanism of adverse effects reduction of electrical stimulation on tissue and electrode, and the methods which could reduce the pain and persecution were investigated in this study. In accordance with the experimental and theoretical evidence, we proposed a new conceptual framework for cyborg mechanism and categorized the types of stimulation based on the different regions of the neural system. We also explored the effective parameters to achieve optimal stimulation. In this regard, we proposed using the following options:

- the capacitance property of the electrode for reducing the Faradaic current.
- series capacitors which can cause the reduction of the Faradaic current, especially in constant voltage of stimulation.
- a charge balance stimulation signal around 100Hz frequency.
- an ultra-high frequency ($\sim KHz$) which is modulated by a low frequency carrier signal ($\sim 100Hz$), which the carrier signal preferably must be applied randomly to reduce the adverse effects of stimulation, to cause successful stimulation in weak intensity cases, increasing the response probability, and reducing the adaptive effect.

In the continue, experimentally, a new method of fish locomotion control was developed. In our method, more attention was made to the stimulation method and neural region of stimulation. The forward locomotion and turning of fish motion were also controlled. However, fundamentally, there are still questions about the controllability mechanism of the creatures and their sense under stimulation, which requires more extensive research.

## Author Contributions

M. J. and Y. J. designed the research, conceived the experiments, and developed the method and theory. The theory and data were analyzed and controlled by M. G. The results were discussed and interpreted by M. J. The manuscript was written and revised by M. G. M. J. and Y. J.

## Additional Information

The authors declare that they have no competing interests.